\documentclass[twocolumn,english,aps,prb,twocolum,superscriptaddress,nobibnotes,amsmath,amssymb,floatfix]{revtex4-2}
\usepackage[colorlinks=true,citecolor=blue,linkcolor=magenta]{hyperref}

\usepackage[markup=nocolor, authormarkupposition=left]{changes} 
\usepackage{soul}
\usepackage[utf8]{inputenc}
\usepackage[english]{babel}
\usepackage{amsmath,amsfonts,amssymb,esint}
\usepackage[T1]{fontenc}
\usepackage{url}

\usepackage{amsmath}
\usepackage{amsfonts}
\usepackage{amssymb}
\usepackage{upgreek}

\usepackage{epstopdf}
\usepackage{graphicx}
\graphicspath{{./Figures/}}

\usepackage{changes}

\begin{document}

\title{A heterogeneously integrated lithium niobate-on-silicon nitride photonic platform}

\author{Mikhail Churaev}
\thanks{These authors contributed equally to this work.}
\affiliation{Institute of Physics, Swiss Federal Institute of Technology Lausanne (EPFL), CH-1015 Lausanne, Switzerland}

\author{Rui Ning Wang}
\thanks{These authors contributed equally to this work.}
\affiliation{Institute of Physics, Swiss Federal Institute of Technology Lausanne (EPFL), CH-1015 Lausanne, Switzerland}

\author{Viacheslav Snigirev}
\thanks{These authors contributed equally to this work.}
\affiliation{Institute of Physics, Swiss Federal Institute of Technology Lausanne (EPFL), CH-1015 Lausanne, Switzerland}

\author{Annina Riedhauser}
\affiliation{IBM Research - Europe, Zurich, Säumerstrasse 4, CH-8803 Rüschlikon, Switzerland}

\author{Terence Bl\'esin}
\affiliation{Institute of Physics, Swiss Federal Institute of Technology Lausanne (EPFL), CH-1015 Lausanne, Switzerland}

\author{Charles M\"ohl}
\affiliation{IBM Research - Europe, Zurich, Säumerstrasse 4, CH-8803 Rüschlikon, Switzerland}

\author{Miles A. Anderson}
\affiliation{Institute of Physics, Swiss Federal Institute of Technology Lausanne (EPFL), CH-1015 Lausanne, Switzerland}

\author{Anat Siddharth}
\affiliation{Institute of Physics, Swiss Federal Institute of Technology Lausanne (EPFL), CH-1015 Lausanne, Switzerland}

\author{Youri Popoff}
\affiliation{IBM Research - Europe, Zurich, Säumerstrasse 4, CH-8803 Rüschlikon, Switzerland}
\affiliation{Integrated Systems Laboratory, Swiss Federal Institute of Technology Zurich (ETH Z\"{u}rich), CH-8092 Z\"{u}rich, Switzerland}

\author{Ute Drechsler}
\affiliation{IBM Research - Europe, Zurich, Säumerstrasse 4, CH-8803 Rüschlikon, Switzerland}

\author{Daniele Caimi}
\affiliation{IBM Research - Europe, Zurich, Säumerstrasse 4, CH-8803 Rüschlikon, Switzerland}

\author{Simon H\"onl}
\affiliation{IBM Research - Europe, Zurich, Säumerstrasse 4, CH-8803 Rüschlikon, Switzerland}

\author{Johann Riemensberger}
\affiliation{Institute of Physics, Swiss Federal Institute of Technology Lausanne (EPFL), CH-1015 Lausanne, Switzerland}

\author{Junqiu Liu}
\affiliation{Institute of Physics, Swiss Federal Institute of Technology Lausanne (EPFL), CH-1015 Lausanne, Switzerland}

\author{Paul Seidler}
\email{pfs@zurich.ibm.com}
\affiliation{IBM Research - Europe, Zurich, Säumerstrasse 4, CH-8803 Rüschlikon, Switzerland}

\author{Tobias J. Kippenberg}
\email{tobias.kippenberg@epfl.ch}
\affiliation{Institute of Physics, Swiss Federal Institute of Technology Lausanne (EPFL), CH-1015 Lausanne, Switzerland}

\maketitle

\textbf{The availability of thin-film lithium niobate on insulator (LNOI) and advances in processing have led to the emergence of fully integrated LiNbO$_3$  electro-optic devices\cite{Zhu:21, Wang2019_2, Desiatov:19, Bahadori:20}, including low-voltage\cite{Wang2018}, high-speed modulators\cite{He2019}, electro-optic frequency combs\cite{Zhang2019}, and microwave-optical transducers \cite{Holzgrafe:20,McKenna:20}.
Yet to date,  LiNbO$_3$ photonic integrated circuits (PICs) have mostly been fabricated using non-standard etching techniques that lack the reproducibility routinely achieved in silicon photonics.
Widespread future application of thin-film LiNbO$_3$ requires a reliable and scalable solution using standard processing and precise lithographic control. 
Here we demonstrate a heterogeneously integrated LiNbO$_3$ photonic platform that overcomes the abovementioned challenges by employing wafer-scale bonding of thin-film LiNbO$_3$ to planarized low-loss silicon nitride (Si$_3$N$_4$) photonic integrated circuits\cite{Liu:21}, a mature foundry-grade integrated photonic platform. 
The resulting devices combine the substantial Pockels effect of LiNbO$_3$ with the scalability, high-yield, and complexity of the underlying Si$_3$N$_4$ PICs.  Importantly, the platform maintains the low propagation loss ($\mathbf{<0.1}$ dB/cm) and efficient fiber-to-chip coupling ($<$2.5 dB per facet) of the Si$_3$N$_4$ waveguides. 
We find that ten transitions between a mode confined in the Si$_3$N$_4$ PIC and the hybrid LiNbO$_3$ mode 
produce less than 0.8 dB additional loss, corresponding to a loss per transition not exceeding 0.1 dB. 
These nearly lossless adiabatic transitions thus link the low-loss passive Si$_3$N$_4$ photonic structures with electro-optic components.
We demonstrate high-Q microresonators, optical splitters, electrically tunable photonic dimers, electro-optic frequency combs, and carrier-envelope phase detection of a femtosecond laser on the same platform, thus providing a reliable and foundry-ready solution to low-loss and complex LiNbO$_3$ integrated photonic circuits.}

\begin{figure*}[t!]
\centering
\includegraphics[width=\textwidth]{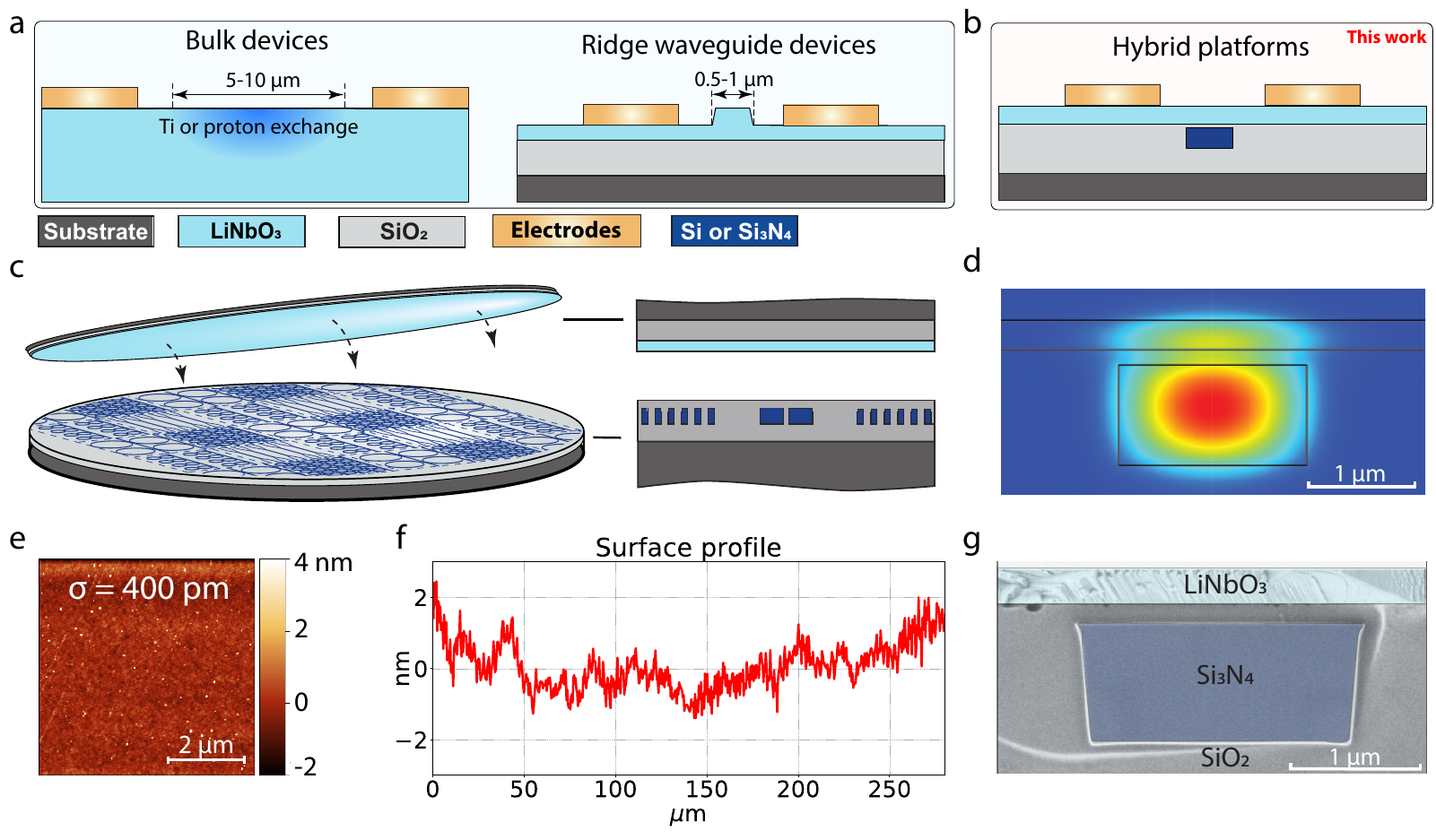}
\caption{ 
\textbf{Hybrid integrated Lithium Niobate photonics}.(a) Conventional approaches to lithium niobate photonics, consisting of traditional Ti or proton exchange based waveguides, and the recently emerged integrated photonics based on etching of thin-film LNOI (which is mostly based on ridge waveguides) (b) The hybrid approach presented in this work, based on heterogeneous integration of thin film lithium niobate with Si$_3$N$_4$. (c) Schematics of our approach which is based on wafer bonding of 4'' (100 mm) thin film lithium niobate onto planarized ultra low loss $\mathrm{Si_3N_4}$ photonic integrated circuits. (d) Hybrid optical mode profile for a typical waveguide used in this work (cf Methods).  (e) AFM measurements of the Si$_3$N$_4$ wafer before bonding showing 400 pm RMS roughness over 5 $\mu$m by 5 $\mu$m field of view. (f) Long-range profilometry scan of the wafer before bonding. (g) False-coloured scanning electron micrograph showing the hybrid structure.} 
\label{Fig:Fig1}
\end{figure*}

Modern society has an constantly increasing demand for optical communications bandwidth, with aggregate data rates doubling every 18 months\cite{Agrell2016,Kitayama2019}, and 
optical technology is coming ever closer to the central processing units\cite{Caulfield2010}. 
Optical modulators play a crucial role in this context, providing the means to transfer electronic signals to optical carriers. 
With the rise of commercial integrated photonics \cite{Thomson:16}, a wide variety of modulation platforms have been demonstrated that are compatible with wafer-scale manufacturing, among which silicon and indium phosphide are the most prominent \cite{Xu2005, Maram2019,Ogiso2017,Witzens2018}. 
In the last decade,  alternative systems, including organic hybrids \cite{Lee2002,Alloatti2014}, plasmonic devices \cite{Haffner2015}, and modulators based on two-dimensional materials \cite{Liu2011,Gruhler:13, Phare:15, Datta:20} have also been developed. 
Among all the materials used, lithium niobate (LiNbO$_3$) remains the most preferable because of its excellent physical properties, and commercial availability \cite{Poberaj2012}. 
Advances in wafer-scale transfer of LiNbO$_3$ thin-films via the SmartCut$^\textbf{TM}$ technique, combined with improvements in etching of LiNbO$_3$, have enabled low-loss integrated electro-optics \cite{krasnokutska2018,Zhang:17,Luke2020}. 
This has led to several key demonstrations, including ultra-high-$Q$ optical microresonators\cite{Zhang:17}, efficient electro-optic frequency comb generation\cite{Zhang2019}, frequency converters \cite{Hu2020}, and non-reciprocal devices \cite{Shao:20, Sohn:21}. 
In addition, electro-optic modulation both at CMOS voltage levels and at high speed (up to 100~GHz) has been achieved \cite{Wang2018,He2019}, offering routes toward compact integrated LiNbO$_3$ modulators compatible with CMOS microelectronics for applications ranging from classical communication for 5G cellular networks and datacenter interconnects to quantum interfaces for microwave-optical conversion \cite{Lambert2020,Javerzac-Galy2016,Wang2019}, and topological photonics employing synthetic dimensions \cite{Yuan2018,Hu2020crystals}. 
Besides the electro-optic applications, integrated LiNbO$_3$ PICs are also of high interest for nonlinear photonics, for example, for efficient second-harmonic generation, optical squeezing, and parametric amplification \cite{Yu2019a,Kanter2002,Lu2021}.

Despite the achievements to date, widespread adoption of LiNbO$_3$ integrated photonics is still impeded by several key issues.  
Current LNOI-based devices are fabricated using specific non-conventional ion-beam etching (IBE) to achieve smooth waveguide surfaces. 
Insufficient etch-mask selectivity leads to the formation of shallow ridge waveguides that require more challenging process control to achieve the desired geometries.
This complicates the establishment of a reliable process design kit (PDK) for integrated LiNbO$_3$ platforms. 
Second, edge coupling between fibers and chips is challenging, as the ridge waveguide structures demonstrated so far show significant coupling loss, 5 to 10~dB per facet, \cite{Hu2020} unless more complicated double-etching techniques are used \cite{He2019edge_coupling,Ying2021}. 
Third, while record resonance quality factors (Q $\approx$ 10$^7$, linear loss of 0.027 dB/cm) have been reported in LiNbO$_3$ ~microresonators\cite{Zhang:17}, this has only been demonstrated for selected optical resonances and has not been achieved broadly in other recently reported works, where losses are typically one order of magnitude higher (0.2 - 0.3 dB/cm, see Supplementary Table 1 for comparison).
For future applications, uniformly low loss across a wafer using precise and mature lithographic processes, along with efficient coupling, are necessary to develop a foundry-level technology that includes PDKs with, e.g., splitters, arrayed-waveguide gratings, optical filters or beamforming networks.

As an alternative to conventional bulk LiNbO$_3$ and ridge-waveguide-based photonic devices, hybrid platforms combining thin-film LiNbO$_3$ with waveguides made of Si, Si$_3$N$_4$, or Ta$_2$O$_5$ have been recently developed\cite{Weigel_2016,Rao:16,Jin:16} (see Fig~\ref{Fig:Fig1}(a)-(c)). 
With proper geometry optimization, the heterogeneously integrated LNOI devices can reach electro-optic performance comparable to that of the all-LNOI platforms\cite{Weigel_2020} 
($\mathrm{V_{\pi} L} = 2.3$ V$\cdot$cm). Heterogeneous integration using the organic adhesive benzocyclobutene (BCB) to bond LNOI to silicon and direct bonding of chiplets to silicon and silicon nitride PICs has been demonstrated, leading to modulators operating at CMOS voltages \cite{Boynton2020,Rao:16}.
Even though the low-loss operation of hybrid Si$_3$N$_4$-LiNbO$_3$ waveguides was reported \cite{Chang:17,Ahmed:19}, the previous works did not provide reliable studies of wafer-scale uniformity, dispersion, insertion loss, or other essential aspects of photonic integrated circuits.
The approaches were aimed at specific device applications and could not demonstrate all the benefits of heterogeneous integration of LiNbO$_3$ with a well-developed photonic platform.
Moreover, the wafer-level bonding required to achieve scalable PDKs was not shown.
Here, we demonstrate a high-yield, low-loss, integrated LiNbO$_3$-Si$_3$N$_4$ photonic platform that solves multiple issues of LNOI integrated photonics. 
The approach circumvents the need for optimized IBE etching of LiNbO$_3$ and opens up the possibility of creating a wide range of low-loss integrated electro-optic photonic circuits. 
This is achieved by wafer-scale heterogeneous integration \cite{Komljenovic2018,Chang:17} (i.e. direct wafer bonding~\cite{plossl1999wafer}) of an LNOI wafer onto a patterned and planarized ultra-low-loss Si$_3$N$_4$ substrate as depicted in Fig~\ref{Fig:Fig1}(c).  
Our approach combines the maturity of Si$_3$N$_4$ integrated photonics with the large Pockels effect of LiNbO$_3$ and enables complex, hybrid PICs that incorporate passive Si$_3$N$_4$ and electro-optic LiNbO$_3$ and exhibit ultra-low  propagation loss (8.5~dB/m).

\begin{figure*}[t]
\centering
\includegraphics[width=\textwidth]{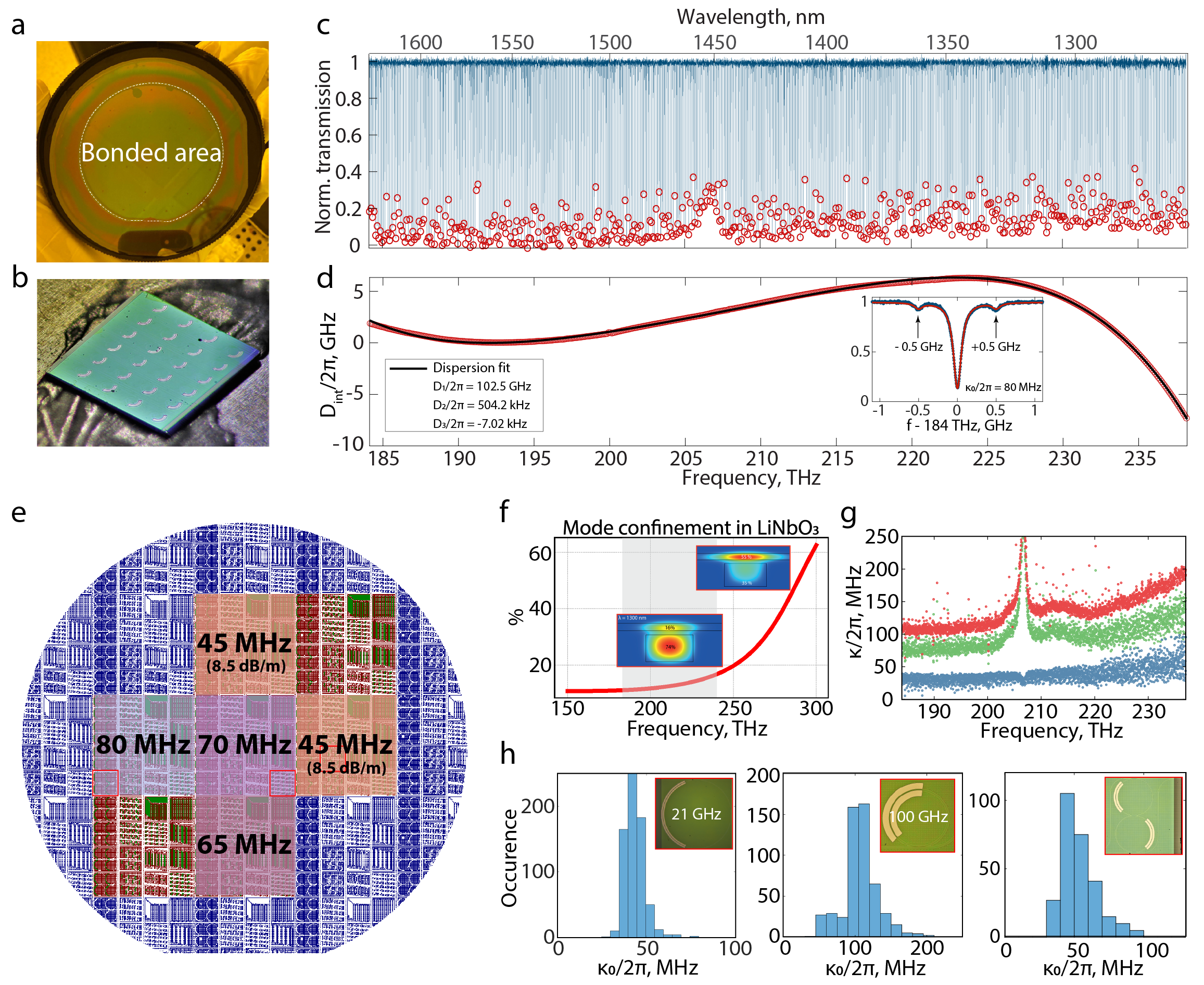}
\caption{
\textbf{Optical loss measurements of the hybrid devices}. (a)  Photograph of a fully bonded 2'' lithium niobate wafer to a 4'' photonic damascene wafer. (b) Schematics of the chip facet with lithium niobate removed from the silicon nitride inverse tapers. FDTD simulations of the mode transition at interface. (c) Broadband transmission measurements of a 100 GHz FSR ring showing flat coupling of resonances from 1260 nm to 1630 nm wavelength. (d) Extracted integrated dispersion of the device. The inset shows a zoom-in of one of the modes at 184 THz with amplitude modulation sidebands at 500 MHz and the corresponding fitting curve. (e) Wafer map with averaged linewidth indicated for the reference 21 GHz FSR rings. (f) Simulated optical mode confinement in LiNbO$_3$ as a function of optical frequency. Insets show typical mode profiles. Grey-shadowed area represents the measurement bandwidth. (g) Loaded (red), intrinsic (green), and coupling (blue) linewidth of the resonances presented in (c). (h) Measured linewidth of 3 types of devices on the wafer accumulated over 55 THz measurement bandwidth.}
\label{Fig:Fig3}
\end{figure*}

\textbf{Fabrication process}. 
The process flow for our hybrid PICs starts with the fabrication of Si$_3$N$_4$ waveguide structures using the photonic Damascene process~\cite{Pfeiffer:18b, Liu:21}.   
We use a 100~mm-diameter silicon wafer with 4~$\mu$m-thick wet thermal SiO$_2$, followed by deep-ultraviolet (DUV) stepper lithography, preform dry etching, preform reflow, low-pressure chemical vapor deposition (LPCVD) deposition of Si$_3$N$_4$, chemical-mechanical polishing (CMP), and SiO$_2$ interlayer deposition and annealing, as detailed in the Supplementary Information. 
The Si$_3$N$_4$ photonic Damascene process is free of crack formation in the highly tensile LPCVD Si$_3$N$_4$ film and provides high fabrication yield and ultra-low propagation loss (1~dB/m).
In addition, double-inverse nanotapers \cite{Liu:18} are incorporated for efficient edge coupling to lensed fibers. 
Previous devices fabricated using this process have been the workhorse for numerous system-level applications of soliton microcombs, ranging from coherent telecommunication \cite{Marin-Palomo:17} to astrophysical spectrometer calibration \cite{Obrzud:19} to supercontinuum generation \cite{Guo:18} and turnkey soliton generation \cite{Shen:20}.  

One of the key advantages of the photonic Damascene process is the possibility of obtaining an extremely flat wafer surface suitable for heterogeneous integration. 
Specifically, we perform CMP on the SiO$_2$ interlayer and bond the fabricated Si$_3$N$_4$ Damascene substrate to a commercially available LNOI wafer (NanoLN).  
The most critical constraints for achieving high bonding yield, the surface roughness and topography, are measured prior to bonding. 
With CMP, the long-range topography is reduced to a few nanometers over several hundred microns, as shown in Fig.  \ref{Fig:Fig1}(f). 
Moreover, atomic force microscopy (AFM) measurements over a range of a few microns reveal a root-mean-square (RMS) roughness of 400 pm 
(Fig. \ref{Fig:Fig1}(e)). 
This roughness level is sufficiently low for direct wafer bonding. 
The donor and the acceptor wafer (the Si$_3$N$_4$ substrate and the LNOI wafer, respectively) are cleaned, and both are coated by atomic layer deposition (ALD) with a few nanometers of alumina (Al$_2$O$_3$). 
The wafers are then bonded and annealed at 250$^\circ$C to enhance the bonding strength. 
See SI for more technical information on fabrication.
We have successfully bonded several wafers, including whole 100-mm LNOI wafers, with a bonding yield close to 100\%, as evidenced by photoacoustic spectroscopy.
Subsequent back-end processing and electrode integration are described in the SI. 
A scanning electron microscope image (Fig. \ref{Fig:Fig1}(g)) of a cross-section of the layer structure reveals clean bonding results. 
Finite-element-method simulations (see Fig.~\ref{Fig:Fig1}(d)) indicate that, for the waveguide and wafer parameters used here (cf. SI for waveguide geometry), the optical mode confinement factor for LiNbO$_3$ at the telecommunication wavelength of 1550 nm is $\Gamma = \iint_{\mathrm{LiNbO_3}}|E|^2dS/\iint_{\mathrm{\Omega}}|E|^2dS = 12\%$, where $\Omega$ denotes the whole cross-section area. 
To demonstrate the electro-optic capabilities of the heterogeneously integrated LiNbO$_3$ photonic circuits, we deposit tungsten ($\mathrm{W}$) electrodes on top of the LiNbO$_3$ adjacent to the waveguides with an electrode-electrode gap of 6~$\mu$m.

To illustrate the versatility, lithographic precision, complexity, and yield of the hybrid platform, we design a reticle with various devices. Figure~\ref{Fig:Fig3}(e) shows the design layout of the Si$_3$N$_4$ photonic circuits for a 100-mm wafer; it contains nine fields with 16 chips each -- in total, more than 100 chips with dimensions of 5 mm $\times$ 5 mm.
The reticle includes chips with three different types of devices: (1) microresonators with a free spectral range (FSR) of either 100 GHz or 21 GHz, the former being used for electro-optic comb generation; (2) photonic molecules consisting of a pair of coupled microresonators each with a FSR of 50 GHz, as used for microwave-optical conversion schemes; and (3) waveguides with a length of several centimeters for supercontinuum generation (wafer D67b01).

\begin{figure*}[t]
\centering
\includegraphics[width=\textwidth]{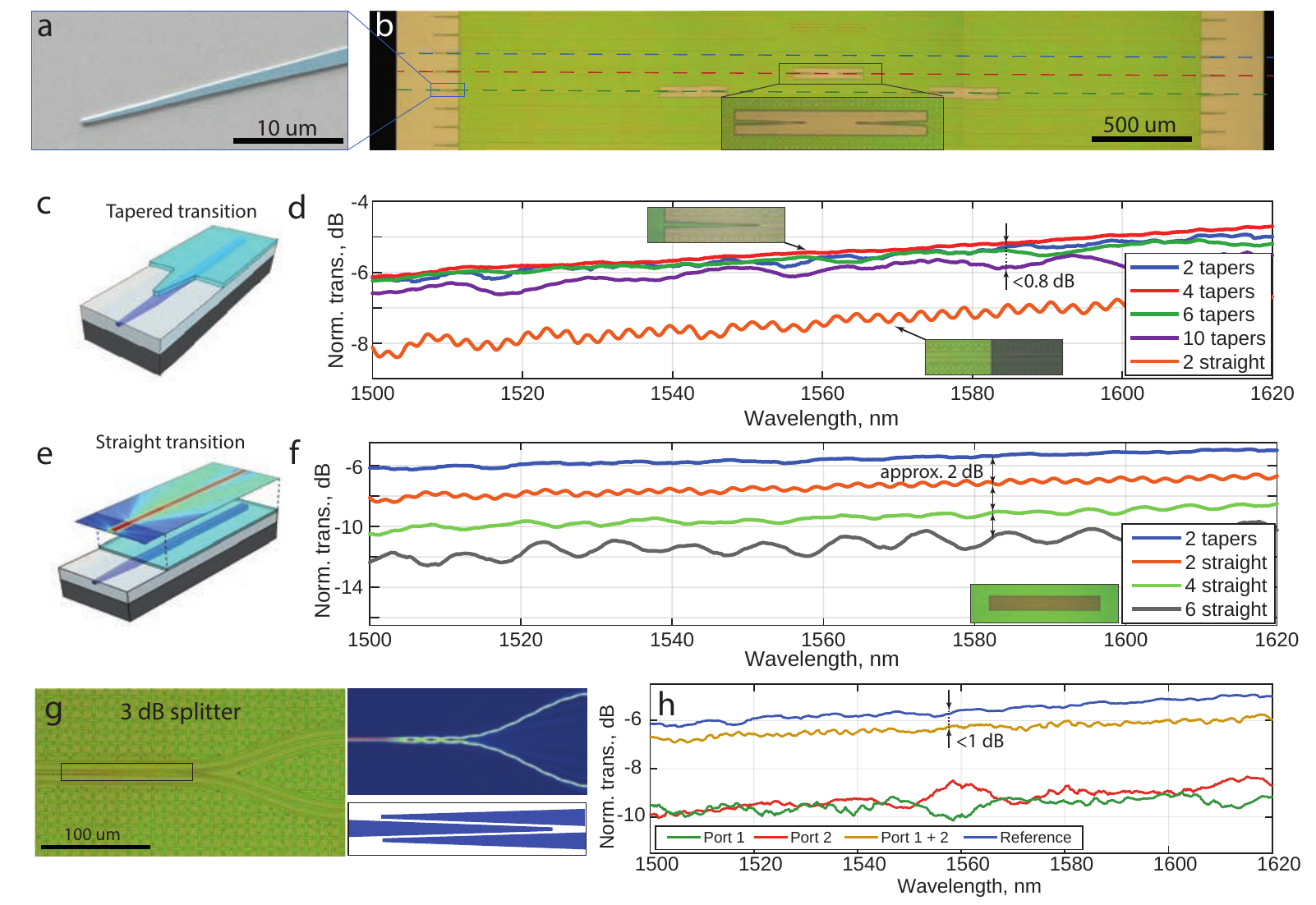}
\caption{\textbf{Adiabatic mode transitions and hybrid optical splitters}. (a) Colored SEM image of a LiNbO$_3$ taper fabricated to make a smooth optical mode transition. (b) Optical microscope image of a chip used to estimate the performance of  interface tapers. Horizontal dotted lines mark waveguides with two interface transitions (blue), four interface transitions (red), and six interface transitions (green). The inset shows a zoom-in of a breakout region. (c) Schematic of a tapered (adiabatic) interface transition. (d) 
Transmission measurements of breakout waveguides described in (b) as well as a typical straight (non-adiabatic) transition for comparison (orange line). (e) Schematic of a straight (non-adiabatic) interface transition. (f) Transmission measurements of waveguides with straight transition breakouts. (g) Optical image, FDTD simulation, and schematic of a W-type 3-dB splitter. The geometry is defined by the underlying Si$_3$N$_4$ layer. (h) Transmission measurements of two ports of the splitter together with the transmission of a straight waveguide as reference.}
\label{Fig:Fig_tapers}
\end{figure*}

\textbf{Optical loss characterization}. 
Linear propagation loss is crucial in determining the performance of integrated electro-optic devices, as it limits the length of electro-optic modulators and the complexity of the associated photonic circuit.
To measure the linear optical loss, evanescent coupling properties, and group velocity dispersion (GVD) of the hybrid structures, we perform broadband frequency-comb-assisted
spectroscopy \cite{Liu2016} of multiple microresonators across the entire wafer with three different external-cavity diode lasers covering the wavelength ranges of 1260–1360 nm, 1355–1505 nm, and 1500–1630 nm. 
The intrinsic quality factors of individual 100~GHz microring resonators reach up to $ Q = 3\times 10^6$, while the 50~GHz photonic dimers (i.e., coupled microrings) and 21~GHz single rings exhibit even higher quality factors up to $Q = 4.5\times 10^6$.
The latter corresponds to a linear propagation loss of 8.5~dB/m.
We observe an absorption peak at approximately 1420~nm (207~THz), which we associate with an overtone of OH-bond vibrations in lithium niobate \cite{Schwesyg2010,Heinemeyer2006}. 
As shown in Fig \ref{Fig:Fig3}(g), optical losses rise with increasing optical frequency. 
We associate this dependency with increased whispering-gallery (radiation) loss as the mode shifts into the LiNbO$_3$ thin film at higher frequencies and becomes less confined (see Fig \ref{Fig:Fig3}(f)). 
For the same reason, we observe nearly uniform evanescent coupling of optical microresonators over a span of 55~THz. 
The layout in Fig \ref{Fig:Fig3}(e) is labeled with the most probable linewidth measured for 21~GHz microresonators in various regions, indicating the degree of variation across the wafer (see Supplementary Information). 
These results not only demonstrate high yield and wafer-scale fabrication but include some of the highest quality factors achieved to date with integrated LiNbO$_3$ devices.
Notably, the $Q$s reported here are not isolated values as in prior work on ridge resonators\cite{Zhang:17} but are consistently high, i.e., we measure hundreds of resonances with $Q$ above $4\times 10^6$  (linewidth below 50~MHz), as shown in Fig. \ref{Fig:Fig3}(h). 

Our hybrid platform also offers the possibility of precise dispersion engineering; due to the interplay between material dispersion and the optical mode distribution, the dispersion can be adjusted by varying the Si$_3$N$_4$ waveguide geometry or the LiNbO$_3$ thickness.
In this work, we designed the structure to work in the near-zero GVD regime advantageous for broadband optical frequency comb generation. 
As shown in Fig.~\ref{Fig:Fig3}(d) for a microresonator with a FSR of 100~GHz, the measured integrated microresonator dispersion ($D_\text{int}$) only varies by 15~GHz over 
an optical bandwidth of 55~THz. 
Interestingly, there is no constraint preventing the design of devices that are uniformly coupled over a broad frequency range and, at the same time, have extremely small dispersion.

\begin{figure*}[t]
\centering
\includegraphics[width=\textwidth]{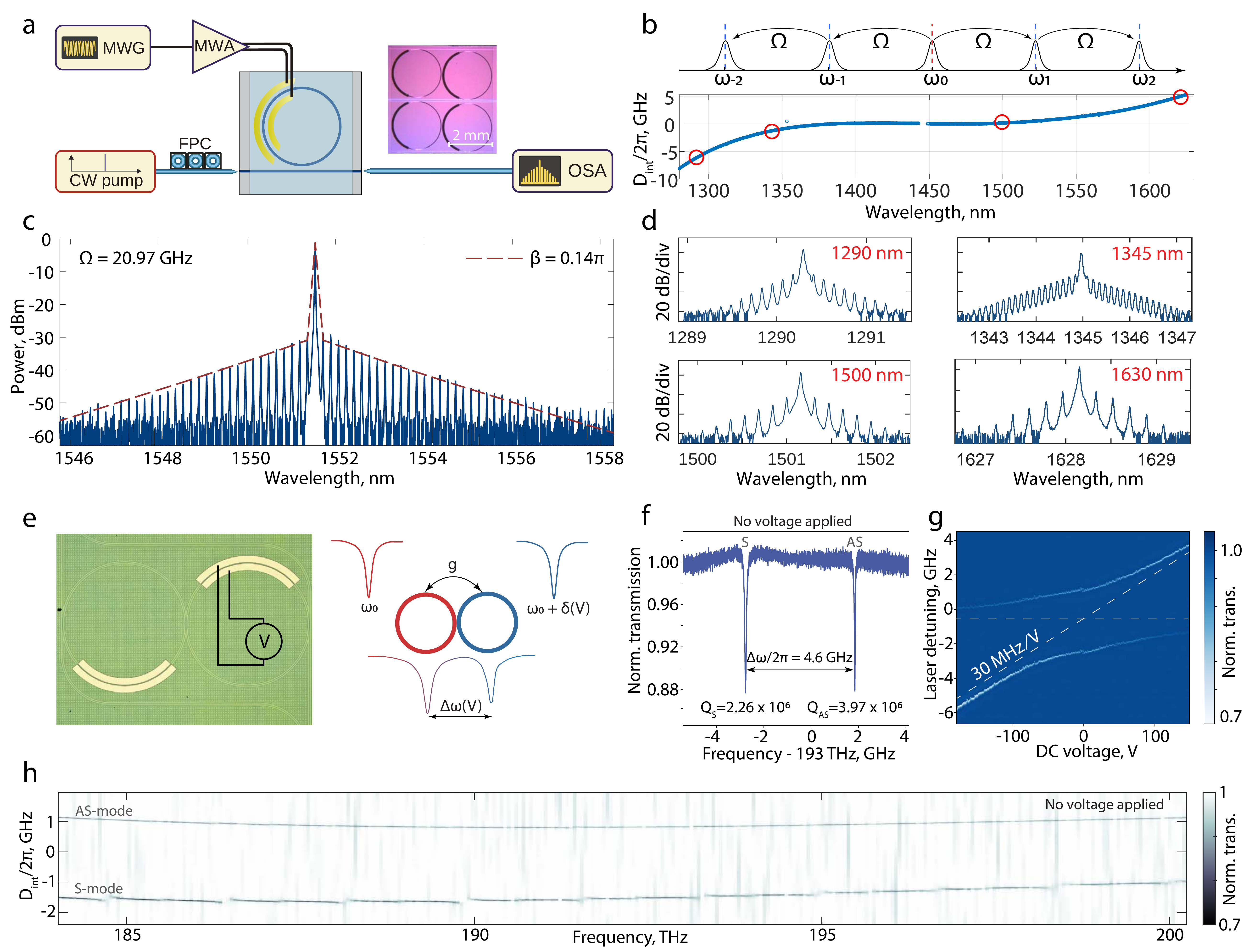}
\caption{
\textbf{Electro-optic frequency comb generation and tunable dimers using heterogenously integrated $\mathrm{LiNbO_3}$ photonic circuits.}  (a) Experimental setup for electro-optic frequency comb generation. MWG - microwave generator, MWA - microwave amplifier, OSA - optical spectrum analyser. (b) Mode coupling schematics and integrated dispersion of the measured device.(c) Measured optical spectrum of the generated EO comb at 1552 nm central frequency. Dashed line corresponds to numerical simulations with phase modulation amplitude $\beta$ = 0.14 $\pi$. (d) Examples of EO frequency combs generated at 4 other pump wavelength. (e) Photonic dimer image and mode hybridization illustration. (f) High-Q resonance splitting of a photonic dimer without additional biasing. S - symmetric supermode, AS - antisymmetric. (g) DC tuning of the photonic dimer mode hybridization, corresponding to linear tuning of around 30 MHz/V for a single mode. (h) Echelle plot of the photonic dimer transmission, showing mode hybridization over a broadband scanning range.}
\label{Fig:Fig4}
\end{figure*}

\textbf{Fiber-chip coupling and interface transitions.} 
Efficient input coupling from an optical fiber to the photonic chip is paramount for numerous applications.
For air-cladded LiNbO$_3$ ridge waveguides, inverse tapers lead to fiber-chip edge-coupling losses of around 10~dB per facet unless complicated multi-layer etching is used\cite{He2019edge_coupling,Mercante2016,Hu:21}. This is due to the significant mode mismatch between the lensed fiber mode (typically circular with about 2.5~$\mu$m diameter) and the asymmetric mode of partially etched air-cladded ridge waveguide structures.
Some recent work on integrated LiNbO$_3$ devices demonstrated the possibility of using embedded silicon edge-couplers to overcome this challenge\cite{He2019}.
In our case, the relative indices of refraction of the materials would lead to significant coupling loss if the LiNbO$_3$ layer remained on top of underlying Si$_3$N$_4$ inverse tapers.
Hence, we remove the LiNbO$_3$~from the coupling regions and rely on standard Damascene Si$_3$N$_4$ inverse tapers\cite{Liu:18}. 
While this provides an efficient input coupling, there remains the challenge of the transition between 
the regions with and without LiNbO$_3$.
To address this issue, we designed and implemented adiabatic tapers in the LiNbO$_3$~layer, as shown in Fig.\ref{Fig:Fig_tapers}(c). The tapers are 100 $\mathrm{\mu}$m long with a tip width of 500 nm and a final width of 10 $\mathrm{\mu}$m.
Both film removal and etching of the tapers are done in a single fabrication step using argon ion-beam etching with a photoresist etch mask (see the Supplementary Information for details on taper fabrication).
All the functional photonic components do not depend on the etching of the chip interface, which is employed only at mode transition regions and, due to the short taper length, does not require low roughness.
We thus keep the LiNbO$_3$~layer unprocessed for all the photonic components, where both roughness and precise alignment are critical to achieving low optical losses.

To measure the efficiency of the adiabatic transitions and remove the ambiguity associated with the fiber-chip coupling loss from the measurement,  we designed an experiment in which we introduce multiple breakouts on straight waveguides, where the optical mode experiences transitions from Si$_3$N$_4$ waveguides into the hybridized mode and back as depicted in Fig.~\ref{Fig:Fig_tapers}(b).  
We fabricated waveguides with 2, 4, 6, and 10 transitions (input/output tapers and 0, 1, 2, and 4 breakouts, respectively) to determine the increase in loss due to each transition.
Figure \ref{Fig:Fig_tapers}(d) shows the results of these measurements. 
As a reference, we compare them with straight interfaces (schematic in Fig.\ref{Fig:Fig_tapers}(e)). 
As can be seen in Figure \ref{Fig:Fig_tapers}(f), each straight transition leads to approximately 1~dB loss, 
whereas the tapered input/output behaves in this measurement as a virtually lossless transition. 
Strikingly, for the case of the tapered transitions, we observe hardly any difference between 2, 4 and 6 interfaces, as shown in Figure \ref{Fig:Fig_tapers}(d) and approximately only 0.8~dB additional loss for ten interfaces.
Considering the statistical uncertainty in the measurements, we deduce a transition loss of $<$0.1~dB per taper.
Calibration of the transmission measurements is discussed in the Supplementary Information.

The lithographic precision of the Si$_3$N$_4$~photonic circuit layer provides our heterogeneous integration approach with versatiliy and robustness, as confirmed by the implementation of a W-shaped 3-dB splitter/coupler\cite{Wang_splitter:16} (see Fig. \ref{Fig:Fig_tapers}(g)) that uses the hybrid Si$_3$N$_4$-LiNbO$_3$ mode but is defined solely by underlying Si$_3$N$_4$ inverse tapers.  Splitters are important components for many optical devices, such as electro-optic modulators, optical networks, and lasers based on reflective semiconductor optical amplifiers.
The elegance of this type of splitter is in its simplicity of design.
Due to the presence of the LiNbO$_3$~slab and the single-mode nature of our hybrid waveguides, the optical mode is adiabatically transferred from the input arm to the output arms.
We make the tapered sections 100 $\mu$m long, 
ensuring a small footprint for integrated components exploiting this design.
Transmission measurements of the device reveal a flat response, with power asymmetry between the two arms not exceeding 1.7 dB and on-chip insertion loss not exceeding 1 dB in the 1500-1620 nm wavelength range (Fig.\ref{Fig:Fig_tapers}(h)).

\textbf{Electro-optic devices.} 
To demonstrate the electro-optic performance achievable with our high-quality factor, hybrid LiNbO$_3$~microresonators, we generate electro-optic frequency combs in the 21~GHz resonators pumped resonantly in the telecommunications C-band. We apply a high-power microwave signal with a frequency 20.97~GHz across the integrated electrodes such that microwave-induced sidebands are resonantly enhanced (Fig \ref{Fig:Fig4}(a)-(b)). 
Even though only 12\% of the optical mode is confined inside the lithium niobate, it is enough to achieve strong modulation of the optical phase. We observe around 60 sidebands within a 25~dB span for an injected RF power of 40~dBm, as depicted in Fig~\ref{Fig:Fig4}(c). 
In this experiment, the phase modulation amplitude corresponds to approximately 0.14$\mathrm{\pi}$ (see Supplementary Information).  The electro-optic coupling is enhanced due to the device's high quality factor and flat dispersion. 
We also make use of the previously mentioned homogeneous coupling of our hybrid devices at optical wavelengths ranging from 1260~nm to 1630~nm to generate electro-optic comb at five different pump wavelengths (1290 nm, 1345 nm, 1500 nm, 1550 nm, and 1625 nm) on a single device (Fig~\ref{Fig:Fig4}(d)).
Moreover, according to the simulations, the geometry can be optimized for maximum electro-optic efficiency with a characteristic $\mathrm{V_{\pi}\cdot L}$ product comparable to (2x larger than) the performance of X-cut ridge-waveguide platforms\cite{Wang2018}. 
The reason is that, in ridge waveguides, most of the electro-optic interaction occurs not in the ridge itself, but in the slab layer, where the modulating electric field is a factor of magnitude stronger (see Supplementary Information). Therefore, the ridge waveguides and hybrid waveguides are conceptually similar in terms of electro-optic interactions (i.e., the role of the slab in ridge waveguides is being taken over by the bonded LiNbO$_3$ layer in our hybrid structure).
Similar studies on optimization of electro-optic performance of heterogeneously integrated LiNbO$_3$ devices can also be found elsewhere\cite{Weigel_2020}.
As a further example of the electro-optic capabilities, we fabricate photonic dimers, which are known building blocks for quantum coherent transducers based on cavity electro-optics\cite{Wade2015,Javerzac-Galy2016,Zhang2019_mol,McKenna:20,Youssefi2021}.
Figure \ref{Fig:Fig4}(g) shows the mode hybridization in the system as a function of the applied DC voltage. 
We observe frequency tuning of 30~MHz/V when a DC voltage is applied to one of the rings.
Moreover, the precise and mature fabrication of the Si$_3$N$_4$ waveguides enables the creation of high-Q photonic dimers exhibiting broadband normal mode splitting even at zero-bias (cf Figure \ref{Fig:Fig4}(f)-(h)). 
The presence of avoided mode crossings for the symmetric supermode (lower frequency) is common with the photonic dimer configuration\cite{Tikan:2022}.
In one last experiment exploiting the $\chi^{(2)}$ nonlinearity of the LiNbO$_3$, we perform supercontinuum generation in the hybrid waveguides.
We observe octave-spanning supercontinuum generation mediated by the $\chi^{(3)}$ nonlinearity, together with simultaneous second-harmonic generation due to the optical field in the LiNbO$_3$, allowing direct measurement of the carrier-envelope offset frequency of the femtosecond pulse laser used as a pump. The details of this experiment can be found in the Supplementary Information.

\textbf{Summary.} 
To conclude, we have demonstrated a hybrid Si$_3$N$_4$-LiNbO$_3$ platform for photonic integrated circuits using direct wafer-scale bonding that endows the mature Si$_3$N$_4$ Damascene technology with the second-order nonlinearity ($\chi^{(2)}$ / Pockels effect) of LiNbO$_3$. 
The heterogeneous integration preserves the precise lithographic control, low propagation loss, and efficient fiber-to-chip coupling of the underlying Si$_3$N$_4$ waveguides for use in a variety of important photonic building blocks. 
We have also presented a design for the transition from Si$_3$N$_4$ waveguides to hybrid Si$_3$N$_4$-LiNbO$_3$ waveguides with a measured insertion loss not exceeding 0.1~dB per interface.
The ability to achieve low-loss transitions is essential for the realization of complex devices, providing a bridge between passive silicon nitride photonics and electro-optic devices.
To the best of our knowledge, this is the first time a heterogeneously integrated LiNbO$_3$ photonic platform combines all the beneficial features of Si$_3$N$_4$~PICs at wafer scale. 
A comparison of the simultaneously achieved desirable features is given in Supplementary Table 1. 
With further geometry optimization, the electro-optic performance can reach levels comparable to that of ridge waveguide structures while keeping propagation losses independent of the quality of the LiNbO$_3$ etching.
Possible applications of our platform include photonic switching networks for neuromorphic or quantum computing, devices for quantum state transduction from microwave to optical photons, integrated electro-optic frequency comb sources, on-chip generation of second-harmonic and squeezed light, as well as high-speed electro-optic devices for optical communications or rapidly tunable, low-noise lasers.
\\

\begin{footnotesize}

\noindent \textbf{Acknowledgments}: 
We thank the Operations Team of the Binnig and Rohrer Nanotechnology Center (BRNC), and especially Diana Davila Pineda and Ronald Grundbacher, for their help and support. Silicon nitride substrates were fabricated in the EPFL center of MicroNanoTechnology (CMi).
We also thank Aleksandr Tusnin for his help in numerical simulations.
\\

\noindent \textbf{Funding Information}: 
This work was supported by funding from the European Union Horizon 2020 Research and Innovation Program under the Marie Skłodowska-Curie grant agreement No. 722923 (OMT) and No. 812818 (MICROCOMB), as well as under the FET-Proactive grant agreement No. 732894 (HOT). This work was also supported by the Swiss National Science Foundation under grant agreement No. 176563 (BRIDGE) and 186364 (Sinergia).  \\

\noindent\textbf{Data Availability}: 
The code and data used to produce the plots are available from Zenodo.

\end{footnotesize}

\bibliographystyle{apsrev4-2}
\bibliography{LNOD_ref}

\end{document}